 \documentclass[preprint2]{aastex}

\newcommand{\til}{$\sim$}
\newcommand{\targ}{X\thinspace0512-401}

\shorttitle{Optical identification of the X-ray burster in NGC1851 }
\shortauthors{Homer et al.}

\begin{document}


\title{OPTICAL IDENTIFICATION OF THE X-RAY BURSTER\\ IN THE GLOBULAR CLUSTER NGC1851\footnote{Based on observations with the NASA/ESA Hubble
Space Telescope, obtained at the Space Telescope Science Institute, which is operated by the Association of Universities for Research in
Astronomy, Inc, under NASA contract NAS5-26555.}}


\author{L. Homer, Scott F. Anderson, Bruce Margon, Eric W. Deutsch}
%\author{Scott F. Anderson}
%\author{B. Margon}
\affil{Astronomy Department, Box 351580, University of Washington,
    Seattle, WA 98195}
\email{homer,anderson,margon,deutsch@astro.washington.edu}
\and
\author{Ronald A. Downes}
\affil{Space Telescope Science Institute, 3700 San Martin Drive, Baltimore, MD 21218}
\email{downes@stsci.edu}
%%\altaffiltext{1}{Email: homer@astro.washington.edu}
%%\altaffiltext{2}{Current address:}

\begin{abstract}
We have obtained exposures of the field of \targ\ in the globular cluster NGC1851, in X-rays with the {\it Chandra X-ray Observatory}, and in the
far-UV with the {\it Hubble Space Telescope}.  We derive an accurate new X-ray position (within \til1\arcsec) for \targ, which enables us to
confirm that the {\em only} plausible candidate for the optical/UV counterpart is the Star~A, which we previously identified from WFPC2 imaging.  We
find no evidence for X-ray or UV flux modulation on the ultra-short ($\la1$~hr) expected binary period, which implies a
low system inclination.  In addition, we have detected and spatially resolved an X-ray burst event, confirming the association of the burster,
quiescent X-ray source, and optical object.  The very large $L_{X}/L_{opt}$ of this object implies an extraordinarily compact system, similar to
the sources in NGC6624 and NGC6712.
\end{abstract}



\section{INTRODUCTION}
The nature of the 12 bright ( $>$10$^{36}$ erg s$^{-1}$) X-ray
sources in globular clusters \citep[see e.g.][]{verb95,bail96} appears distinct from that of low-mass X-ray binaries (LMXBs) in the
Galaxy as a whole. That they are LMXBs was established by mass estimates based on their cluster positions, as measured by the {\it Einstein}/HRI \citep{grin84}.  However, they are over-abundant by a factor $\sim$100  requiring
entirely different formation mechanisms \citep[see][]{verb88}.  Moreover, studies of the limited number of optical counterparts \citep{deut98T,deut00} imply that their period distribution is also
remarkably different from that of field LMXBs, with a preponderance of ultra-short period systems.  NGC\,6624
harbors the shortest period binary system known (with $P\simeq 11$min),
and our {\it HST} observations of the X-ray source in NGC\,6712 \citep{home96} indicated that it too is likely to be a similarly exotic system.  Indeed, the only companion to the neutron star primary that can fit into such a compact binary is a white dwarf, making them double degenerates-- remarkable endpoints to binary stellar evolution.

Even given the 3\arcsec\ (90\% confidence limit) {\it Einstein} X-ray positions, optical identifications are very difficult, due to the extremely crowded locations.  In the case of NGC\,1851, \citet{deut98} WFPC2 imaging revealed
\til300 stars within the {\it Einstein} error circle.  Their proposed counterpart, Star~A, is a very
strong candidate, given its similarity in color (faintness and large UV-excess) to the confirmed counterpart in NGC6712, yet a \til5\% {\it a posteriori}
probability still remained that such a UV-excess star could coincidentally lie within the {\it Einstein} X-ray error circle. Furthermore,
earlier ground-based work by \cite{auri94} suggested that another UV-bright object with colors of a horizontal branch star, X1, might be an
unusual counterpart. Clearly, additional observations were needed to confirm or refute these various suggested identifications.

To that end, we have obtained both {\it Chandra}/HRC data of the field in order to significantly improve the X-ray position, and a set of time resolved {\it HST}/STIS FUV-images to search for any variability.  We
present the results of these observations in this Letter.

\section{OBSERVATIONS AND DATA REDUCTION}
\subsection{\it Chandra X-ray Observatory}
{\it Chandra} observed the field of \targ\ for 12ks on 1999 December 25.  The high resolution
camera + low-energy
transmission grating spectrograph mode \citep[HRC-S+ LETG;][]{murr97,brin97,pred97} we chose provides the highest possible spatial resolution
available from {\it Chandra}, to achieve our primary science goal, but also a high resolution low energy spectrum of the bright source.  The
spectral results will be presented elsewhere, whilst we will concentrate on the positional result here.  

Data reduction was initially undertaken with routines in {\tt CIAO v1.1.5}. The anti-coincidence shield of HRC-S is not operational, owing to
a timing error in the electronics. This leads to a much higher background rate of (false) events.  However, the intrinsic energy resolution of the detector (though poor) can be used to easily remove 25\% of this by excluding the
highest energy channel. A sliding-cell detection routine ({\tt tgdetect}) confirmed that only the one source was strongly detected, the LMXB, and also
centroided its position to within 0.03\arcsec\ (0.2 pix).  A lightcurve was also extracted using all available data.  Regions were defined covering the 0th and
1st order images, and 4 rectangular background regions above and below.  The events were then summed into 2s bins, the background
scaled and lastly
subtracted using the {\tt CIAO} routine {\tt lightcurve}.
 
\subsection{\it Hubble Space Telescope}
Four orbits (\til 12 ks) of {\it HST}/STIS \citep{kimb98}  imaging were undertaken on 1999 March 24.  We used the FUV MAMA with the long-pass quartz filter in
time-tag mode, to provide complete flexibility for temporal analysis. Its $25\arcsec\times25\arcsec$ field of view covered both the cluster core and the entire {\it
Einstein} X-ray error circle.  In this  \til1400\AA--1700\AA\ passband even the core of the globular
cluster is uncrowded (see fig.~\ref{fig:FUVchart}, lower left), and we found that 0.2\arcsec\ radius aperture photometry worked best
(optimized according to a curve of growth analysis).
%Indeed, an attempt at PSF fitting using {\tt DAOPHOT} \citep{stet87} revealed significant systematic morphology changes during each orbit, which had undesirable affects.
To limit systematic effects we applied differential photometry, whereby the magnitudes of the stars of interest were calculated relative to
an ensemble of the brightest stars in the field. 

\section{REFINING THE X-RAY POSITION}
\begin{figure}[!htb]
\plotone{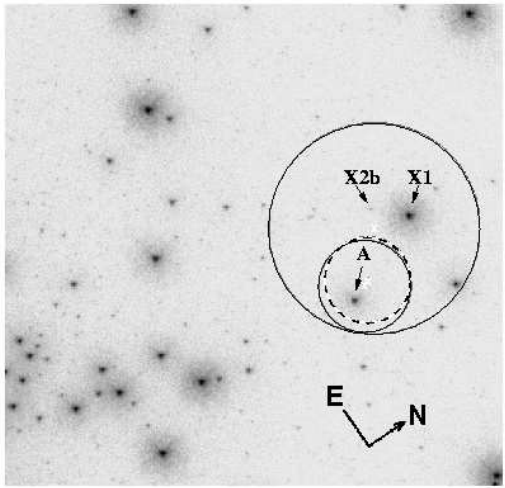}
\caption{Deep (13ks) {\it HST}/STIS FUV image of the core of NGC1851.  The large error circle (3\arcsec\ radius) is that of the {\it Einstein}/HRI X-ray
position.  The two smaller circles indicate the major refinement afforded by {\it Chandra}/HRC-S, and show that using either the Tycho-2
(solid) or USNO A-2 (dashed) catalogs to calculate the offset to the {\it HST} Guide star catalog makes no significant difference: both include Star~A, whilst excluding all other UV-bright objects.  All
circles represent 90\% confidence. \label{fig:FUVchart}}
\end{figure}
Although the superlative PSF of {\it Chandra} enables a very precise determination of a relative position for the X-ray source, we require an
absolute value.  Unfortunately, the lack of other bright X-ray sources in the field means that we cannot derive a precise absolute position by
correcting relative to known positions.  Instead, we rely on the calibration of the observatory's aspect.  The  {\it Chandra} team has made a
detailed study of the aspect behaviour, based on comparisons between the X-ray positions of all sources with identifications and precise positions in the
optical/radio, specifically objects appearing in the Hipparcos/Tycho-2 \citep{perr97,hog00}, USNO A-2 \citep{mone98} and ICRF \citep{ma98}
catalogs.  This has revealed a long-term drift in the aspect of each of the detector
systems, although only 10 data points are available for HRC-S observations to date. We were provided with the appropriate corrections (from a
linear fit) for the
date of our observation.  For all detectors combined the residuals then have an rms of 0.6\arcsec,
whilst for the HRC-S subset the value is slightly larger, 0.7\arcsec.  We adopt this latter value as an estimate of the intrinsic aspect
uncertainty.
\begin{table*}
%\small
\caption{X-ray and optical/UV positions in the ICRS.\label{tab:offsets}}
\begin{center}
\begin{tabular}{l l l  l l c l l} 
\tableline\tableline
{\small Source of }&{\small No. of  }&{\small $\Delta$(RA)}&{\small $\Delta$(DEC)} &\multicolumn{2}{c}{{\small Source positions}}&{\small Dataset}\\
{\small Positions }&{\small matches }&{\small (seconds) }&{\small (arcsecs)} &{\small RA(2000)}& {\small DEC(2000)}&\\
\tableline
{\em Chandra} Aspect			&		&					& &5:14:6.43 & -40:02:37.63&{\em Chandra} X-ray\\
USNO A-2 & 267 & $-0.02\pm0.01$ &$-0.16\pm0.13$ & 5:14:6.41 & -40:02:38.22&{\em HST} optical/UV \\
Tycho-2  & 36 & $-0.01\pm0.02$ & $-0.33\pm0.33$ &5:14:6.42 & -40:02:38.05&{\em HST} optical/UV\\
\tableline
\end{tabular}
\end{center}
\end{table*} 

Before we can overlay our X-ray position onto the {\it HST}/FUV image, a final correction must be made for different frames of reference.
Essentially the FUV image can be tied directly to the $B$ band images used by \cite{deut98}, upon which accurate astrometry was performed to
derive the optical positions of the various stars of interest within the {\it Einstein} error circle\footnote{We are able to confirm the identity of the
FUV star with Star~A on the basis of its spectral energy distribution.  The MAMA flux measure of 8.7$\pm0.4\times10^{-16}$erg\,cm$^{-2}$\,s$^{-1}$
is fully consistent with the previous FOS and WFPC2 results.}.  However, this astrometry made use of the then current
{\it HST} Guide star catalog \citep[HST-GSC;][]{jenk90}, which predates the adoption of the International Celestial Reference System \citep[ICRS;][]{feis97}, in which the
Hipparcos/Tycho, USNO A-2  and ICRF catalog are grounded.  Hence, we have calculated the offsets between HST-GSC and ICRS by matching stars
appearing in both HST-GSC and either Tycho-2 or USNO A-2 within 6\arcmin--30\arcmin\ of the position of \targ\ (thereby excluding the crowded
regions at the globular cluster center).  The results are given in Table~\ref{tab:offsets}.  For clarity we combine the rms uncertainty in
this final frame offset with the {\it Chandra} positional uncertainty, yielding 90\% error radii of 1.3\arcsec\ and 1.2\arcsec\ for Tycho-2 and USNO A-2 offsetting
respectively.  In both cases, the {\it Chandra} X-ray position area is reduced by a factor of 6 from the {\it Einstein} result, and as shown in
figure~\ref{fig:FUVchart}, the \cite{deut98} candidate Star~A still lies squarely within.  We can also now finally exclude X1 \citep{auri94} at the 99.9\%
confidence level.  Star~A is in fact the {\em only} UV bright object visible within the {\it Chandra} error circle, and the probability of a
chance alignment has been reduced to $\la1\%$ following the arguments of \cite{deut98}.

\section{X-RAY/FUV VARIABILITY}
Globular cluster X-ray sources are thought to essentially all be X-ray
bursters. Despite the very high quiescent X-ray flux of this source, there
appears to be only one previously-published X-ray burst, observed a
quarter-century ago from {\it Uhuru} \citep{form76}, obviously
with very limited angular resolution. Our {\it Chandra} observations
fortuitously observe and spatially resolve an X-ray burst (Figure 2). We
thus can now definitively associate the quiescent X-ray source X0512-401,
the X-ray burster, and the optical counterpart.

We have also searched for variability in the persistent X-ray emission, but find none, setting a 99\% confidence upper limit of 4\% on the semi-amplitude of any periodic
modulation between 1~min and 1.5~hr (after subtraction of a 385s sinusoid, an artifact of dithering).
\begin{figure}[!htb]
\plotone{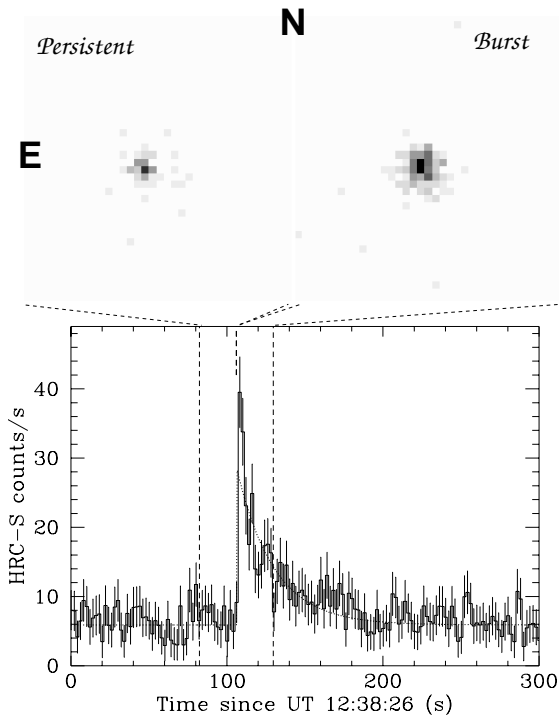}
\caption{Upper panel: \targ\ in NGC1851 imaged by {\it Chandra}/HRC-S immediately prior to and during the peak of an X-ray burst event. Each image is a 23.6s
exposure corresponding to one e-folding time of the exponential burst decay. Lower panel: Lightcurve detail with 2~s binning showing the flux evolution of the burst. The fit to a fast-rise exponential decay model
is over-plotted.  The decay time is consistent with the only other published result, from {\it Uhuru}. \label{fig:burst}}
\end{figure}

	Interestingly, our results on the FUV variability are equally null. As expected Star X1 appears to be a constant source (upper limit of 2.4\%), but
so does Star~A.  The limit on Star~A of 5\% on the semi-amplitude flux modulation between 5~min and 6~hr provides a significant
constraint.  For comparison, the measured UV modulation semi-amplitudes for the sources in NGC6624 and NGC6712 are \til8\% and \til4\% respectively \citep{ande97,home96}.
The assumption that any UV modulation would arise from the varying contribution of the 
X-ray heated donor star's face implies a low inclination  ($\la30\degr$) for the NGC1851 system.  Lastly, we also checked for any burst events in
the reprocessed UV flux, but again we found nothing.
 
\section{CONCLUSIONS} 
Our {\it Chandra} observations confirm the bursting nature of, and provide a new precise and accurate X-ray position for, the luminous source \targ\ in the
core of NGC1851. Our {\it Chandra} error circle is \til6$\times$ smaller than the previous one from {\it Einstein}.  Comparison to a deep {\it HST}/STIS FUV image shows that the \til1\arcsec\ error circle now excludes all
other FUV bright stars, hence providing yet stronger support for our previous identification of the faint (M$_{\rm B}$=5.6), UV-excess
(\ub=-0.9) Star~A as the optical/UV counterpart to \targ.  STIS time-tag data were searched for variability on timescales from
5~min to 6~hr, but none was found, requiring a low system inclination.

The extremely low optical luminosity of Star~A (confirmed here to be the only UV bright object in the accurate {\it Chandra} error circle), resulting in a relatively high X-ray to optical luminosity ratio,
implies an ultra-compact system, where the small accretion disc provides relatively little reprocessing area. Both the similarity of the
optical/UV spectral
energy distributions of the counterpart in NGC6712 and Star~A \citep{deut98T,deut00}, and the broad-band X-ray spectra of these sources and that in
NGC6624 \citep{sido00}, further support the premise that the NGC1851 system should have
a ultra-short binary period, $\la$1~hr. 

%We conclude that, with this confirmation of Star~A, the evidence that the LMXBs in globular clusters represent a distinct, exotic class of ultra-compact systems is growing, yet its apparent low inclination prevents any direct measurement of its orbital period.
\vspace{2mm}
%\section{Acknowledgments}
We thank Tom Aldcroft of the {\it Chandra} Aspect Team for assistance with the calibration of the X-ray position and estimates of its
uncertainty. Financial support for this work was provided through NASA grant NAG5-7932, STScI grant GO-07363.01-96A and SAO grant GO0-011x.

\end{document}